\documentclass[10pt,twocolumn,letterpaper]{article}

\usepackage[utf8]{inputenc}
\usepackage{amsmath,amssymb,amsfonts}
\usepackage{cite}
\usepackage{graphicx}
\usepackage{algorithm}
\usepackage{algorithmic}
\usepackage{textcomp}
\usepackage{xcolor}
\usepackage{listings}
\usepackage{subcaption}
\usepackage[accsupp]{axessibility}  % Improves PDF readability for those with disabilities.
\usepackage{wacv}              % To produce the CAMERA-READY version
\definecolor{wacvblue}{rgb}{0.21,0.49,0.74}
\usepackage[pagebackref,breaklinks,colorlinks,allcolors=wacvblue]{hyperref}

\def\wacvPaperID{19} % *** Enter the WACV Paper ID here
\def\confName{WACV}
\def\confYear{2026}

\title{Transforming Video Subjective Testing with Training, Engagement, and Real-Time Feedback\\
}

\author{
Kumar Rahul, Sriram Sethuraman, Andrew Segall, Yixu Chen \\
Amazon Prime Video
}

\begin{document}
\maketitle
\begin{abstract}
Subjective video quality assessment is crucial for optimizing streaming and compression, yet traditional protocols face limitations in capturing nuanced perceptual differences and ensuring reliable user input. We propose an integrated framework that enhances rater training, enforces attention through real-time scoring, and streamlines pairwise comparisons to recover quality scores with fewer comparisons. Participants first undergo an automated training quiz to learn key video quality indicators (e.g., compression artifacts) and verify their readiness. During the test, a real-time attention scoring mechanism, using “golden” video pairs, monitors and reinforces rater focus by applying penalties for lapses. An efficient chain-based pairwise comparison procedure is then employed, yielding quality scores in Just-Objectionable-Differences (JOD) units. Experiments comparing three groups (no training, training without feedback, and training with feedback) with 80 participants demonstrate that training-quiz significantly improves data quality in terms of golden unit accuracy and reduces tie rate, while real-time feedback further improves data quality and yields the most monotonic quality ratings. The new training, quiz, testing with feedback, 3-phase approach can significantly reduce the non-monotonic cases on the high quality part of the R-Q curve where normal viewer typically prefer the slightly compressed less-grainy content and help train a better objective video quality metric.
\end{abstract}

\section{Introduction}
Subjective video quality assessment (VQA) is critical for validating video compression algorithms, streaming technologies, and overall Quality of Experience (QoE)~\cite{ITU_P910,VQEG}. Unlike objective metrics that rely purely on computational models, subjective VQA directly captures human perception, making it the gold standard for validating perceptual video quality~\cite{Mantiuk2012,Ramanarayanan2007}. However, traditional subjective testing methods, such as Absolute Category Rating (ACR) and Degradation Category Rating (DCR), often exhibit significant inter-observer variability and scale bias due to fixed rating scales~\cite{ITU_P910}. Full pairwise comparisons provide greater sensitivity by eliminating scale bias, as users merely choose the better of two videos. Yet, they become impractical for large sets of encodes due to the quadratic growth of required comparisons~\cite{Mantiuk2017}.

Subjective assessments can be severely compromised by participant inattentiveness, misunderstanding of the task, or inconsistent decision-making, particularly in unsupervised or crowdsourced settings~\cite{Figuerola2013,Sakic2014}. Prior studies highlight the substantial impact that observer variability and inadequate training can have on data reliability~\cite{Burmania2016, Egger2015}. Platforms such as Netflix' \textit{e2nest}~\cite{E2NEST} and AVRateNG~\cite{AVRateNG, goering2021voyager} have streamlined large-scale subjective assessments but typically lack integrated mechanisms to ensure real-time participant engagement and adequate training.

To address these challenges, recent research emphasizes the importance of comprehensive participant training and real-time attentiveness monitoring~\cite{Orduna2023, Mirkovic2014}. Structured training sessions prior to assessment have shown significant improvements in reliability~\cite{Kittel2023}, while real-time attention monitoring demonstrates measurable improvements in rating consistency and reduced data rejection rates~\cite{Figuerola2013}. Šakić et al.~\cite{Sakic2014} quantified the detrimental impact of inattentive responses on crowdsourced subjective quality scores, further underscoring the necessity of integrated real-time monitoring.

Motivated by these insights, we propose a comprehensive framework integrating three core innovations: (1) an automated participant training quiz with immediate feedback to ensure competency in identifying video artifacts, (2) a real-time attention scoring system leveraging strategically selected ``golden pairs'' to continuously monitor engagement, and (3) an efficient chain-based pairwise comparison protocol that reduces complexity from $\mathcal{O}(n^2)$ to $\mathcal{O}(n)$ while recovering quality scores in Just-Objectionable-Difference (JOD) units~\cite{Mantiuk2017}. Recent studies further emphasize the importance of tailored VQA strategies; for instance, Zhu et al.~\cite{yixu_icme_2025} demonstrated the challenges of resolution cross-over in live sports, while other works~\cite{rahul_icip_2022, rahul_icip_2023, rahul_acm_2023} offer insights on MOS recovery, parameter-driven modeling, and the impact of test environments.

Our controlled study with 80 participants divided into three groups demonstrates that training significantly improves attention scores compared to untrained participants, while the addition of real-time feedback further reduces tie rates and enhances rating decisiveness. By actively maintaining data quality through training and real-time monitoring, our framework enables reliable subjective testing even with limited participant pools, making it highly suitable for industrial deployments. The following sections present related work, detail our methodology, and evaluate its effectiveness in our controlled user study.

\section{Related Work}
\subsection{Advancements in Subjective Video Quality Assessment}
Traditional subjective assessment methods, such as Absolute Category Rating (ACR) and Degradation Category Rating (DCR), have long been standard for evaluating video quality. However, these methods often suffer from significant inter-observer variability and scale biases due to fixed rating scales~\cite{ITU_P910}. Pairwise comparison approaches, in contrast, have gained attention due to their simplicity and intuitive nature, effectively reducing inter-observer variability~\cite{Mantiuk2012,Ramanarayanan2007}. Recent work by Perez-Ortiz and Mantiuk~\cite{Mantiuk2017} demonstrated that maximum likelihood estimation-based scaling methods could effectively convert pairwise judgments into continuous quality scales, expressed in Just-Objectionable Differences (JODs). Their method significantly reduces the required number of comparisons (from $\mathcal{O}(n^2)$ to $\mathcal{O}(n)$), addressing complexity and cognitive load issues inherent in traditional full-matrix comparison methods. Mohammadi et al.~\cite{Mohammadi2022} further evaluated various sampling algorithms for efficient pairwise subjective assessment, demonstrating that strategic pair selection can maintain assessment quality while reducing participant burden.

\subsection{Participant Training and Guidance}
Observer variability remains a substantial challenge in subjective quality assessments. Targeted training sessions illustrating common compression artifacts have been shown to improve participant consistency and reliability~\cite{Kittel2023}. However, many existing assessment platforms offer minimal or no training, leaving participants unprepared to identify nuanced quality differences. Our proposed methodology integrates an automated training quiz with immediate feedback to familiarize participants with typical artifacts (e.g., blockiness, blurring, ringing, grain loss) and assesses their readiness using a rolling accuracy metric. This structured preparation significantly reduces random or inconsistent responses, enhancing data quality and overall reliability of subjective assessments.

\subsection{Attention Checks and Real-Time Monitoring}
Ensuring participant engagement is critical, especially in unsupervised or crowdsourced environments prone to distractions. As highlighted in \cite{Figuerola2013, Sakic2014}, ensuring data quality in crowdsourced subjective video quality assessments requires robust quality control measures to mitigate the effects of participant inattentiveness. Platforms like Netflix's \textit{e2nest}~\cite{E2NEST} and \textit{AVRateNG}~\cite{AVRateNG} provide scalable subjective testing but often lack built-in real-time attentiveness monitoring. Naderi and Cutler~\cite{Naderi2024} recently demonstrated the effectiveness of crowdsourcing approaches for video quality assessment, while Jenadeleh et al.~\cite{Jenadeleh2024_feedback} showed that providing real-time feedback during crowdsourced visual quality assessment with paired comparisons significantly improves data reliability. Their work on crowdsourced estimation of Just Noticeable Differences~\cite{Jenadeleh2024_jnd} further validates the potential of well-designed feedback mechanisms in maintaining participant attention. In contrast to these approaches that primarily focus on crowdsourcing contexts, our method employs a dynamic real-time attention scoring system based on strategically placed golden pairs in a controlled laboratory setting. By continuously updating attention scores, we actively manage participant engagement throughout the study, minimizing unreliable ratings and improving overall test integrity.

\subsection{Scalability and Integration in Industrial Setups}
Scalability remains crucial when deploying subjective testing in industrial contexts. Platforms such as \textit{e2nest}~\cite{E2NEST}, \textit{AVRateNG}~\cite{AVRateNG}, and the QualityCrowd framework~\cite{Keimel2012} support large-scale evaluations but often assume controlled testing environments and lack integrated participant training or live attentiveness feedback. Our solution employs Amazon Fire TV devices for consistent video playback quality across diverse participant environments. This integrated, end-to-end design ensures robust data collection even with fewer participants, making it highly suitable for industrial deployments. Consequently, our approach contributes significantly to the development of reliable Video Quality Metrics (VQMs) tailored for practical internal use.

Our key contributions build upon and extend these established foundations through: (1) an automated training quiz with rolling score qualification and immediate feedback for competency verification, (2) a dynamic golden pair identification mechanism that adapts based on response consensus to enable real-time attention monitoring with participant-visible scoring, and (3) unified integration of these components with efficient chain-based pairwise comparisons, validated through controlled experiments and deployed at scale (27,000+ ratings across 560 sessions).

\section{Proposed Methodology}
Our system integrates three core components to improve subjective video quality testing: an Automated Training Quiz, a Real-Time Attention Scoring Mechanism, and an Efficient Pairwise Comparison protocol for quality score recovery. In our approach, video playback is organized into pairs, and after each pair is shown, participants are presented with three response options. They are asked to indicate whether the first video is better, the second video is better, or if both videos look similar. This three-choice design is intended to capture both clear preferences and cases where quality differences are imperceptible. 

While the inclusion of a tie option (i.e., "both videos look similar") enables the analysis of the distribution of responses and helps identify video pairs where quality is nearly indistinguishable, it also introduces the risk of spamming. In scenarios where participants are fatigued, bored, or insufficiently trained, the tie option might be overused, potentially diluting the accuracy of the quality scores. To mitigate this risk, our framework is complemented by an automated training quiz that ensures participants are well-prepared to identify critical video artifacts, and by a real-time attention scoring mechanism that continuously monitors participant engagement throughout the test. These integrated components work together to improve the reliability of the subjective assessments while reducing the likelihood of data contamination due to inattentive or indiscriminate responses.

\subsection{Automated Training Quiz}
Prior to the main test, each participant completes a training quiz designed to both familiarize them with typical compression artifacts (e.g., blockiness, blurring, ringing, grain loss) and verify their readiness for the main assessment. In this quiz, participants are presented with a series of curated video pair comparisons. These pairs are specifically selected so that the videos differ significantly in quality and bitrate, while maintaining consistent encoding settings aside from the constant rate factor (CRF), vbv-maxrate, and vbv-buffsize parameters—which significantly influence the overall bitrate in x265 \cite{x265}, an implementation of the HEVC \cite{HEVC} standard. For each video pair, participants choose one of three response options: they assign a value of \( -1 \) if they select the first video as better, \( +1 \) if they select the second video as better, and \( 0 \) if they judge both videos to be of equal quality.

\paragraph{Mismatch Category Definitions.}
To assess participant performance during the training quiz, we define three response categories based on how well the participant's selection aligns with the expected outcome (determined by bitrate and encoding quality differences):
\begin{itemize}
    \item \textbf{Perfect Match}: The participant's choice correctly identifies the higher-quality video (i.e., the video with higher bitrate and better encoding parameters).
    \item \textbf{Close Mismatch}: The participant selects the tie option when there is a clear quality difference, indicating uncertainty where a definitive judgment was expected.
    \item \textbf{Complete Mismatch}: The participant incorrectly identifies the lower-quality video as better, representing a fundamental misunderstanding of the quality difference.
\end{itemize}

After each response, the server sends immediate, customized feedback, including technical details of the video pair such as bitrate and resolution differences, and prompts the participant to review the pair again before retrying. This self-paced, automated training and quiz minimizes random guessing and ensures a baseline level of competency. 

\paragraph{Quiz Mechanics.}
Before the quiz, participants view a 1-minute introductory video demonstrating common compression artifacts (blockiness, banding, grain loss) to set expectations. During the quiz, participants may replay pairs as needed, though they are encouraged to complete within 5 minutes (realistically limiting replays to approximately twice per pair). The rolling score is computed over a trailing window of 10 pairs (including the current pair). Participants can qualify after a minimum of 6 pairs if their rolling score exceeds 60\%. If the score remains below 60\% after 20 pairs, the session terminates. This ensures only competent participants proceed to the main assessment.

To quantify performance, let \( s_i \) denote the score obtained on the \( i \)-th comparison. We adopt the following scoring function:

\begin{equation}
\begin{aligned}
s_i =
\begin{cases}
1.0, & \text{if Perfect Match (correct choice)}; \\
0.25, & \text{if Close Mismatch (nearly correct)}; \\
0, & \text{if Complete Mismatch (incorrect)}.
\end{cases}
\end{aligned}
\end{equation}

This scoring mechanism ensures that participants not only learn to differentiate among the three possible responses but also demonstrates their capability to accurately discern subtle quality differences before proceeding to the main test.

A sliding window of the most recent \( W \) pairs (typically \( W=10 \)) is used to compute a rolling score:
\[
S_{\text{roll}} = \left( \frac{1}{W} \sum_{i=k-W+1}^{k} s_i \right) \times 100\%.
\]
Participants must achieve a rolling score above a preset threshold which was set at 60\% to proceed. 

\subsection{Real-Time Attention Scoring Mechanism}
During the main test, our system continuously monitors participant attentiveness via a real-time scoring mechanism based on \textit{golden pairs}—video pairs for which the expected outcome is known or highly probable. The internal raw attention score is initialized to 100 and updated after each golden pair, while the score displayed to participants is bounded between 0 and 100.

\paragraph{Penalty/Bonus Function}

The attention score update is computed as:
\[
A_{\text{new}} = \max \Big\{ 0, \min \Big\{ 100,\, A_{\text{old}} - \Delta + B \Big\} \Big\},
\]
where \(\Delta\) is the penalty term and \(B\) is the bonus term. In our implementation, the same base function is used for both penalty and bonus calculations; however, the scaling factor differs between them. Specifically, when a participant provides an incorrect response, the penalty is computed based on the number of mistakes, while a bonus is computed based on the number of consecutive correct responses. These counters are managed dynamically such that if a mismatch occurs, the consecutive correct counter is reset and the mistake count is incremented; conversely, when a correct response is given, the mistake count is decremented (or reset if negative) and the consecutive correct counter is incremented.

The penalty function is defined as:
\[
\Delta = 1.0 + 0.4 \cdot (\text{mistake\_count} - 1),
\]
and the bonus function is defined as:
\[
B = 1.0 + 0.2 \cdot (\text{consecutive\_correct} - 1).
\]
In practice, these functions ensure that repeated mistakes incur higher penalties, while a streak of correct responses yields a bonus that partially offsets previous errors. The updated attention score is clamped between 0 and 100 for display purposes, even though the internal raw score (which starts from 100) remains unbounded. This dynamic, dual-mode approach enables robust real-time monitoring of participant attentiveness.

\paragraph{Golden Pair Placement Strategy.}
To ensure accurate evaluation and guide user scoring, golden pairs are introduced in a strategic manner. Initially, pairs with the same resolution and encoder settings—but with a bitrate gap selected according to the training pair selection logic—are predefined as seed golden pairs, which help identify inattentive participants early in the study. Subsequently, after collecting approximately 20 ratings per pair, the system dynamically adjusts the golden pair set based on response ambiguity.

\paragraph{Dynamic Golden Pair Evolution.}
After collecting approximately 20 ratings per pair, we compute the mean preference score $\bar{r}$ and standard deviation $\sigma_r$ for each pair. A pair is promoted to golden status if: (1) $|\bar{r}| > 0.5$ (strong consensus on winner), (2) $\sigma_r < 0.3$ (low variability across raters), and (3) at least 75\% of responses agree on the same winner. Pairs with ambiguous ratings ($|\bar{r}| < 0.5$ or $\sigma_r > 0.5$) are excluded from golden pair consideration to avoid penalizing legitimate perceptual uncertainty. Over time, the increasing number of golden pairs strengthens the scoring system and provides better insights into user attentiveness.

\subsection{Efficient Pairwise Comparison for Compression Optimization}
\begin{figure}[htbp]
    \centering
    \includegraphics[width=0.95\linewidth]{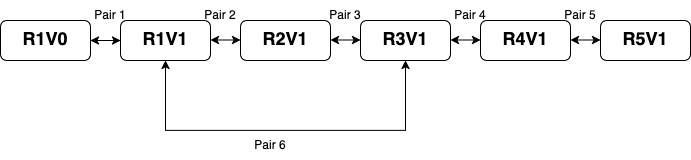}
    \caption{Pairwise comparison strategy.}
    \label{fig:pair_strategy}
\end{figure}

A key novelty of our methodology lies in reducing the total number of pairwise comparisons from \(\mathcal{O}(n^2)\) to \(\mathcal{O}(n)\). Rather than exhaustively comparing all possible encodes, we organize them into a linear chain based on resolution and encoder settings, thereby requiring each new version to be compared only with its immediate neighbor in the quality ladder. As illustrated in Figure~\ref{fig:pair_strategy}, each encode is labeled \(\text{R}i\text{V}j\), where \(\text{R}i\) indicates the resolution (with \(\text{R}1\) being the highest) and \(\text{V}j\) denotes a variant produced by adjusting the constant rate factor (CRF), maxrate, or bitrate in x265.

This chain-based approach not only streamlines compression optimization but also integrates seamlessly with our attention scoring mechanism. Pairwise comparisons that exhibit strong consensus can be promoted to \emph{golden pairs}, thereby reinforcing the scoring system and further improving the detection of inattentive or random responses.

For instance, \(\text{R1V0}\) serves as a pseudo-reference stream representing top quality, followed by lower-quality rungs such as \(\text{R1V1}\), \(\text{R2V1}\), \(\text{R3V1}\), and so on. By chaining these versions sequentially (\(\text{R1V0} \rightarrow \text{R1V1} \rightarrow \text{R2V1} \rightarrow \text{R3V1} \rightarrow \dots\)), we only need \(n-1\) comparisons for \(n\) encodes. This strategy greatly reduces test complexity and cognitive load for participants compared to the \(\binom{n}{2}\) comparisons required by a full pairwise design.

Each comparison produces an empirical preference probability:
\[
\hat{p}_{ij} = \frac{c_{ij}}{n_{ij}}, \quad i \neq j,
\]
where \(c_{ij}\) is the number of times video \(i\) was preferred over \(j\) and \(n_{ij}\) is the total number of comparisons between them.

\paragraph{Tie Response Handling.}
Note that since our model supports only binary responses (i.e., either the first video is better, corresponding to \(-1\), or the second video is better, corresponding to \(+1\)), any tie responses (i.e., a value of 0) are assumed to be equally distributed between the two outcomes when generating the pairwise comparison matrix. This approach is justified because: (1) ties represent perceptual uncertainty rather than a distinct quality judgment, (2) the Thurstone Case V model assumes a continuous quality scale where ties occur when $|q_i - q_j| < $ JND, and (3) equal splitting preserves the symmetry property of the PCM ($p_{ij} + p_{ji} = 1$)~\cite{Mantiuk2017}. In future work, we plan to develop a refined score recovery model that explicitly leverages tie statistics and the final set of golden pairs to better capture perceptual uncertainty.

We then map these probabilities to a Just-Objectionable Difference (JOD) score using the inverse cumulative normal function (probit)~\cite{Mantiuk2017}. Under the Thurstone Case V assumption of equal variance, quality scores \(\{q_i\}\) are recovered via a Maximum Likelihood Estimation (MLE) approach. Specifically, the likelihood of observing \(c_{ij}\) wins out of \(n_{ij}\) comparisons between videos \(i\) and \(j\) is modeled as:
\begin{equation}
\begin{aligned}
L(q_i - q_j \mid c_{ij}, n_{ij}) &= \binom{n_{ij}}{c_{ij}} 
\left[ \Phi\left(\frac{q_i - q_j}{\sigma}\right) \right]^{c_{ij}} \\
&\quad \times \left[ 1 - \Phi\left(\frac{q_i - q_j}{\sigma}\right) \right]^{n_{ij}-c_{ij}},
\end{aligned}
\end{equation}

The quality scores \(\hat{q}\) are obtained by maximizing the overall log-likelihood of the collected data, with \(\sigma\) set such that 1 JOD corresponds to a detection probability of 0.75. Unanimous responses (where \(\hat{p}_{ij}\) is 0 or 1) are adjusted by shifting them to the nearest non-unanimous value before applying the inverse cumulative transformation~\cite{Mantiuk2017}. Notably, this calibration approximately equates 1 JOD to 1 JND, thereby linking our recovered quality scores to a well-established psychophysical threshold.

\section{Experimental Setup}
We conducted a controlled study with 80 participants divided into three distinct groups to investigate the effects of training and real-time attention feedback on subjective video quality ratings. In \emph{Group A (No Training, No Feedback)}, 25 participants performed pairwise comparisons independently, receiving neither training nor real-time feedback. In \emph{Group B (Training, Passive Attention Monitoring)}, participants completed a training session and had to pass a qualification quiz prior to the main study. Their attention was monitored server-side but not communicated to participants; initially 27 participants enrolled, but 2 were disqualified during training, resulting in 25 valid participants. \emph{Group C (Training, Active Attention Feedback)} also required participants to pass the training quiz, but additionally provided real-time attention scores during the main test. Golden pairs for real-time feedback in Group C were dynamically selected based on combined data from Groups B and C. Of 28 participants who began in Group C, one failed the training quiz, leaving 27 participants with valid ratings.

Subjective tests were conducted in a controlled laboratory environment using four OLED displays (two LG2023 OLED55C3PUA and two LG2019 OLED55C9PUA units) supporting native HDR10 playback. All displays were configured in filmmaker mode, which disables post-processing (motion smoothing, dynamic contrast, edge sharpening, noise reduction) to ensure participants evaluated actual encoded quality without display-induced artifacts. The LG2023 units served as calibration benchmarks for the LG2019 units, ensuring consistent luminance and color reproduction. Ambient lighting was maintained at 150 cd/m\(^2\) with viewing distance at 1.5× display height per HDR evaluation guidelines.

Moreover, we selected 10 UHD-HDR mezzanine videos from the Amazon Prime Video catalog. Each video was encoded into six HEVC \cite{HEVC} variants using x265 \cite{x265} by varying key parameters such as the constant rate factor (CRF), vbv-maxrate, and vbv-bufsize, while keeping all other encoder settings constant. One variant per source (designated \(\text{R1V0}\) in Figure~\ref{fig:pair_strategy}) served as a pseudo-reference, encoded at 100 Mbps with CRF = 4 to represent near-reference quality suitable for TV playback. Table~\ref{tab:encoding_params} summarizes the encoding configurations for all six variants.

\begin{table}[htbp]
    \caption{Encoding Parameters (CVBR Mode, vbv-bufsize=2×maxrate)}
    \label{tab:encoding_params}
    \centering
    \footnotesize
    \begin{tabular}{|l|c|c|}
    \hline
    Variant & Resolution & maxrate (kbps) \\
    \hline
    R1V0 & 2160p & 100000 \\
    R1V1 & 2160p & 20000 \\
    R2V1 & 1440p & 12000 \\
    R3V1 & 1080p & 5000 \\
    R4V1 & 720p & 1800 \\
    R5V1 & 480p & 600 \\
    \hline
    \end{tabular}
\end{table}

For each source video, we created six pairs following the chain-based strategy illustrated in Fig.~\ref{fig:pair_strategy}, resulting in 60 pairs in total. These pairs were presented in random order. Pairs corresponding to \(\text{R1V1}\) (comparing R1V0 vs R1V1 with an 80 Mbps maxrate difference, approximately 5× bitrate gap) and \(\text{R3V1}\) (comparing R1V0 vs R3V1 with a 95 Mbps maxrate difference and resolution downgrade, approximately 20× bitrate gap) exhibited substantial and reliably detectable quality differences, making them ideal candidates for initial golden pairs to provide a baseline for attention scoring. Consequently, the experiment consisted of 50 normal pairs and 10 initial golden pairs. Dynamic golden pair selection was restricted to trained participants (Groups B and C). Collected metrics included attention scores, tie rates, replay counts, quality scores recovered in Just-Objectionable-Difference (JOD) units, and total test duration.

This experimental design enabled rigorous evaluation of the efficacy of training, real-time feedback mechanisms, and the efficiency of the minimal, chain-based pairwise comparison strategy for accurate and reliable compression optimization.

\section{Results and Analysis}

\begin{figure}[htbp]
    \centering
    \includegraphics[width=0.95\linewidth]{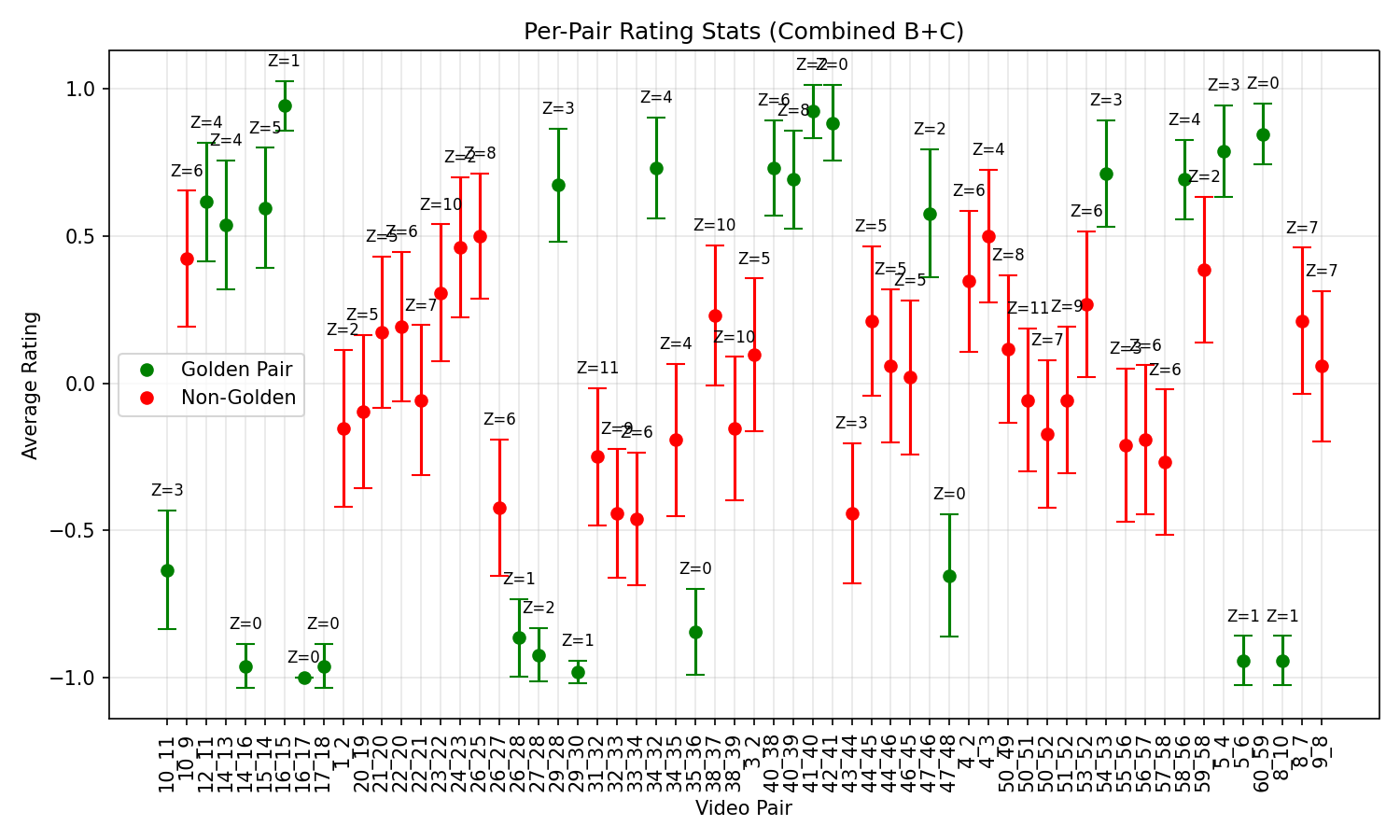}
    \caption{%
    Dynamically selected golden pairs (green markers) from Groups B and C based on response consensus ($|\bar{r}| > 0.5$, $\sigma_r < 0.3$, and $\geq$75\% agreement). Non-golden pairs are shown as red markers.%
    }
    \label{fig:golden-pair-dynamic}
\end{figure}

Figure~\ref{fig:golden-pair-dynamic} shows dynamically selected golden pairs for Groups B and C, serving as attention benchmarks with reliably detectable quality differences.

\begin{figure}[htbp]
    \centering
    \includegraphics[width=0.95\linewidth]{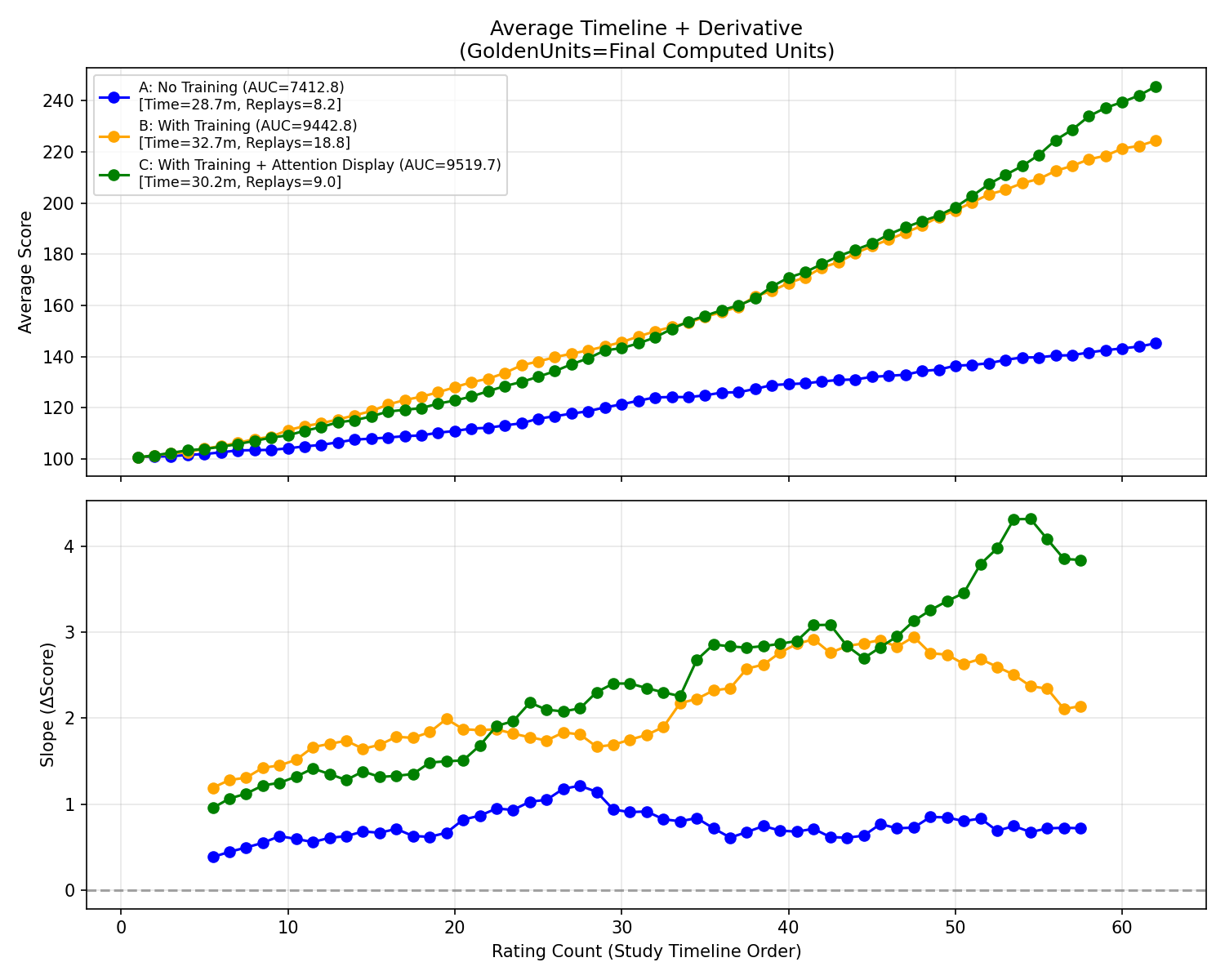}
     \caption{
    Average attention scores (top) and derivative (bottom) across 60 rated pairs. Trained groups (B and C) significantly outperformed Group A ($p < 0.001$). Group C showed significantly lower tie rates than Groups A and B ($p < 0.001$ and $p < 0.01$).}
    \label{fig:avg-attention-derivative}
\end{figure}

Figure~\ref{fig:avg-attention-derivative} demonstrates clear attention trend differences across groups. Trained groups (B and C) showed enhanced scores versus untrained Group A. Group C, with real-time feedback, achieved the highest scores and steepest learning trajectory. The derivative plot shows Group C maintained focus throughout while Group A's attentiveness declined after test midpoint. Real-time feedback did not significantly extend study duration. Table~\ref{tab:summary_metrics} indicates fewer ties in trained groups, showing increased confidence in quality judgments.

\begin{table}[htbp]
    \caption{Summary of Group-Level Metrics (Mean $\pm$ 95\% CI)}
    \label{tab:summary_metrics}
    \centering
    \footnotesize
    \begin{tabular}{|c|c|c|c|c|}
    \hline
    Group & Attn. Avg & Study Time & Replay & Tie Rate \\
          & ($\pm$CI) & (min, $\pm$CI) & ($\pm$CI) & (\%, $\pm$CI) \\
    \hline
    A & 145 $\pm$ 8 & 29 $\pm$ 2 & 8 $\pm$ 6 & 25 $\pm$ 7 \\
    B & 224 $\pm$ 24 & 33 $\pm$ 4 & 19 $\pm$ 10 & 8 $\pm$ 6 \\
    C & 246 $\pm$ 33 & 30 $\pm$ 2 & 9 $\pm$ 5 & 3 $\pm$ 2 \\
    \hline
    \end{tabular}
\end{table}

Post-quiz interviews with the three participants who did not qualify (two from Group~B, one from Group~C) revealed that grain loss artifacts were particularly challenging to identify despite the introductory training video. Participants reported difficulty distinguishing quality differences in grainy clips even after viewing artifact demonstrations. Interestingly, some of these participants re-volunteered in subsequent studies and successfully passed the training quiz with high attention scores, suggesting that initial failures may have resulted from fatigue or that repeated exposure enhanced their perceptual learning.

To assess the statistical significance of group differences, we performed one-way ANOVA tests on key metrics. For attention scores, we found a significant effect of group ($F(2,74) = 47.1$, $p < 0.001$). Post-hoc independent t-tests confirmed that both trained groups significantly outperformed Group~A (Group B vs A: $p < 0.001$, Cohen's $d = -3.01$; Group C vs A: $p < 0.001$, Cohen's $d = -2.47$), while Group~C and Group~B showed no significant difference in final attention scores ($p = 0.292$, Cohen's $d = -0.30$). Similarly, tie rates exhibited significant group differences ($F(2,74) = 56.4$, $p < 0.001$), with Group~C showing significantly lower tie rates than both Group~A ($p < 0.001$, Cohen's $d = 2.99$) and Group~B ($p < 0.01$, Cohen's $d = 0.77$).

Table~\ref{tab:summary_metrics} shows Group~C achieved 157\% higher attention scores than Group~A ($p < 0.001$) with 9\% improvement over Group~B ($p = 0.292$, ns). Most notably, Group~C's tie rate decreased 89\% versus Group~A ($p < 0.001$) and 64\% versus Group~B ($p < 0.01$), demonstrating that while training improves attention, real-time feedback primarily enhances rating decisiveness.

\begin{figure*}[htbp]
    \centering
    % First row: JOD score plots
    \subfloat[Group A: JOD Scores]{%
        \includegraphics[width=0.3\linewidth, trim=50 30 50 50, clip]{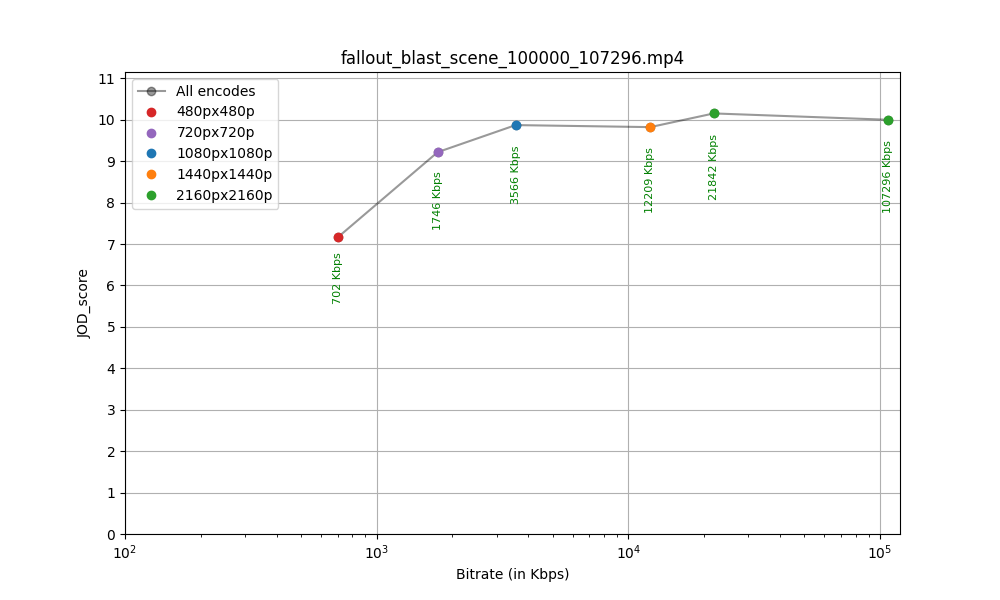}%
        \label{fig:falloutA_JOD}%
    }%
    \hfill
    \subfloat[Group B: JOD Scores]{%
        \includegraphics[width=0.3\linewidth, trim=50 30 50 50, clip]{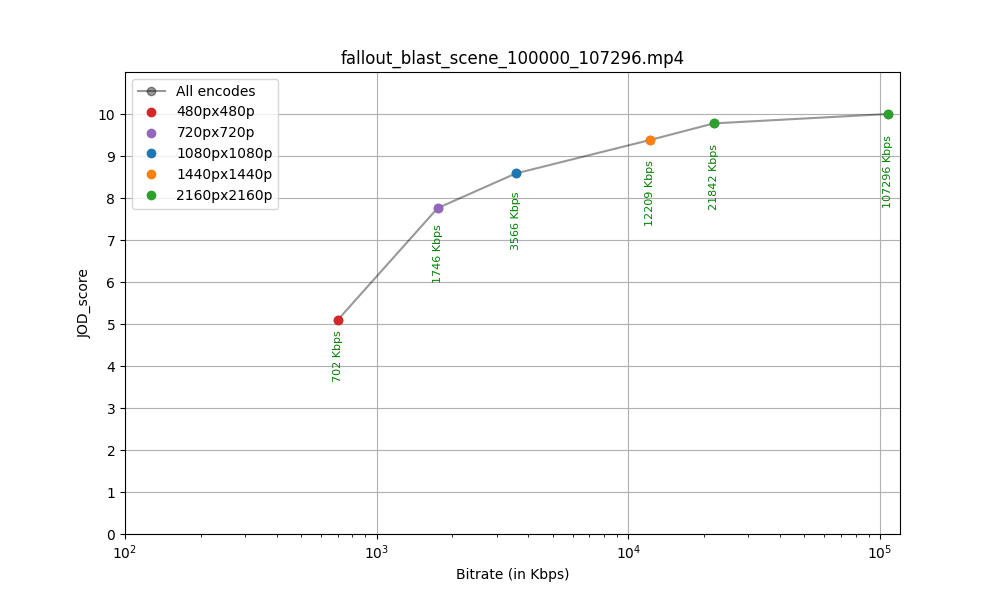}%
        \label{fig:falloutB_JOD}%
    }%
    \hfill
    \subfloat[Group C: JOD Scores]{%
        \includegraphics[width=0.3\linewidth, trim=50 30 50 50, clip]{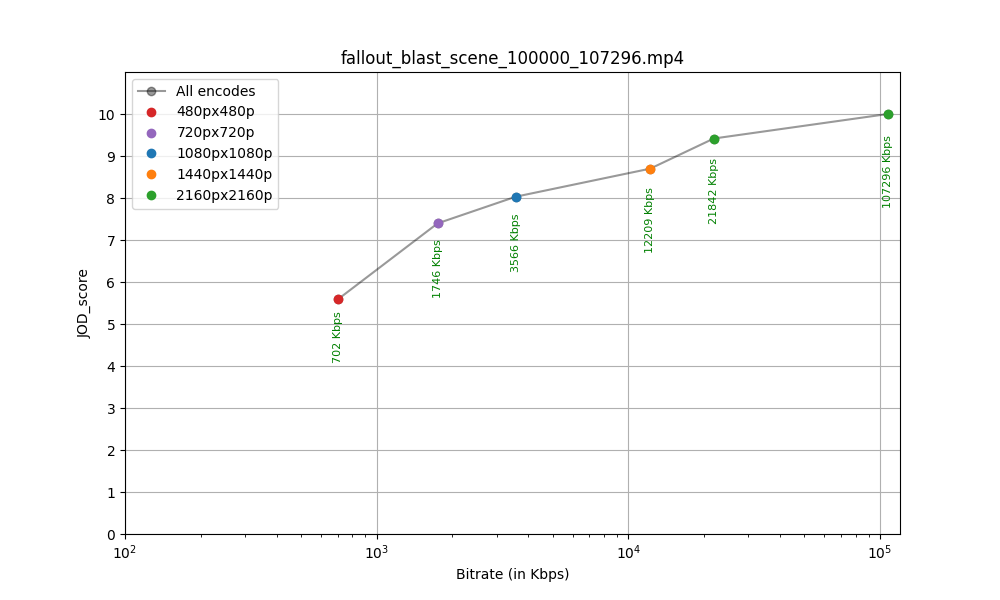}%
        \label{fig:falloutC_JOD}%
    }\\[1ex]
    % Second row: PCM distributions
    \subfloat[Group A: PCM]{%
        \includegraphics[width=0.3\linewidth, trim=50 100 50 140, clip]{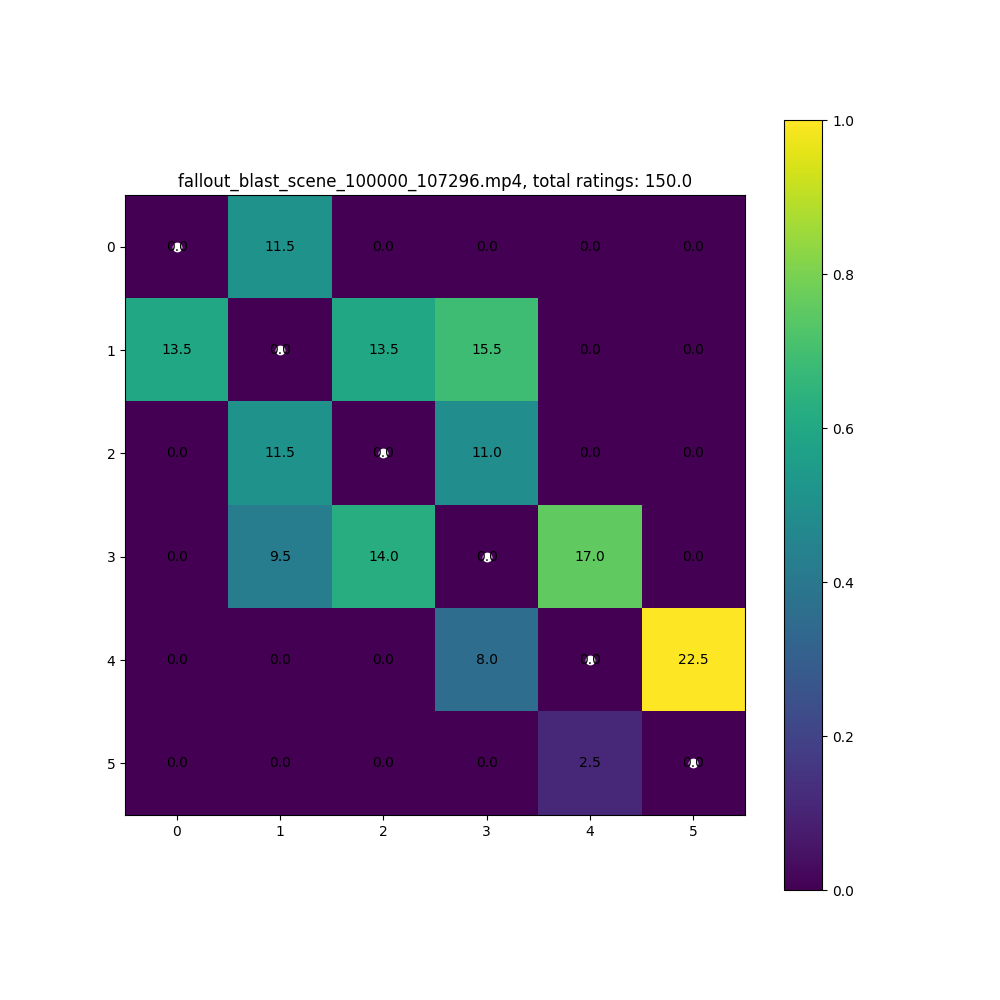}%
        \label{fig:falloutA_PCM}%
    }%
    \hfill
    \subfloat[Group B: PCM]{%
        \includegraphics[width=0.3\linewidth, trim=50 100 50 140, clip]{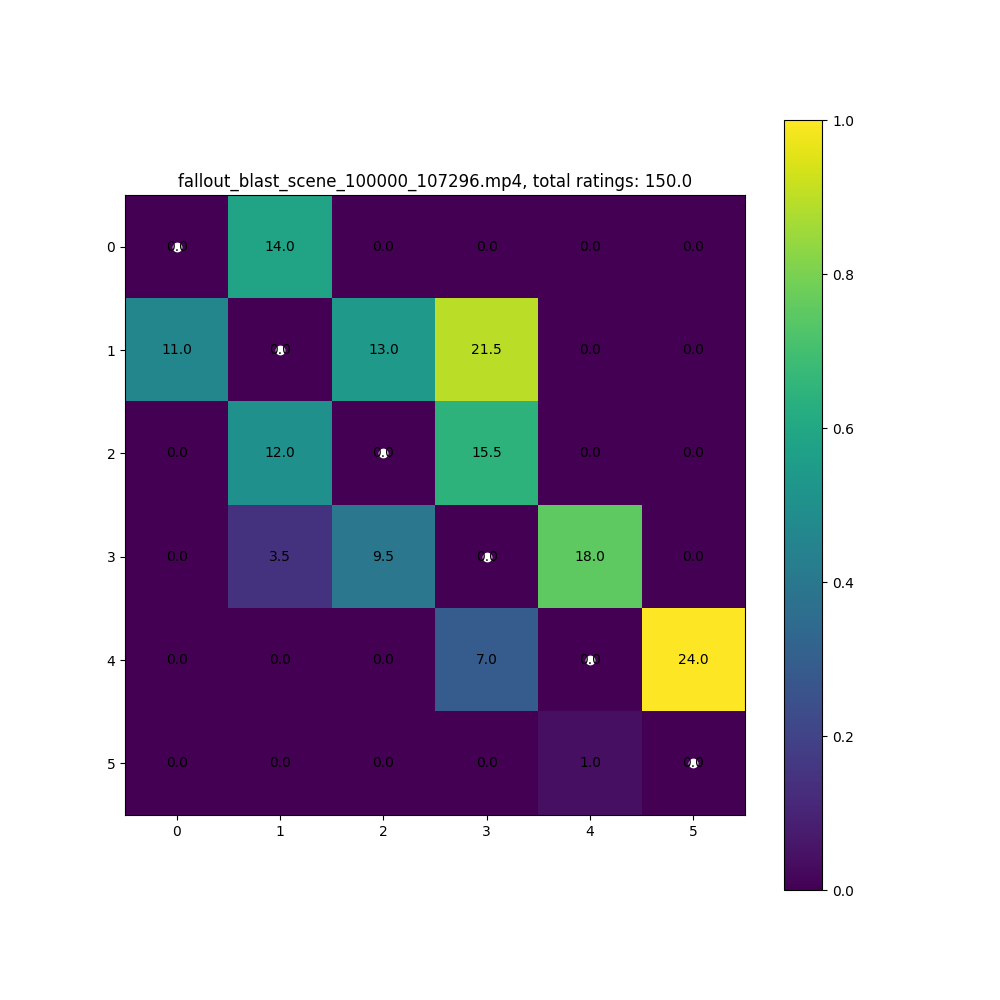}%
        \label{fig:falloutB_PCM}%
    }%
    \hfill
    \subfloat[Group C: PCM]{%
        \includegraphics[width=0.3\linewidth, trim=50 100 50 140, clip]{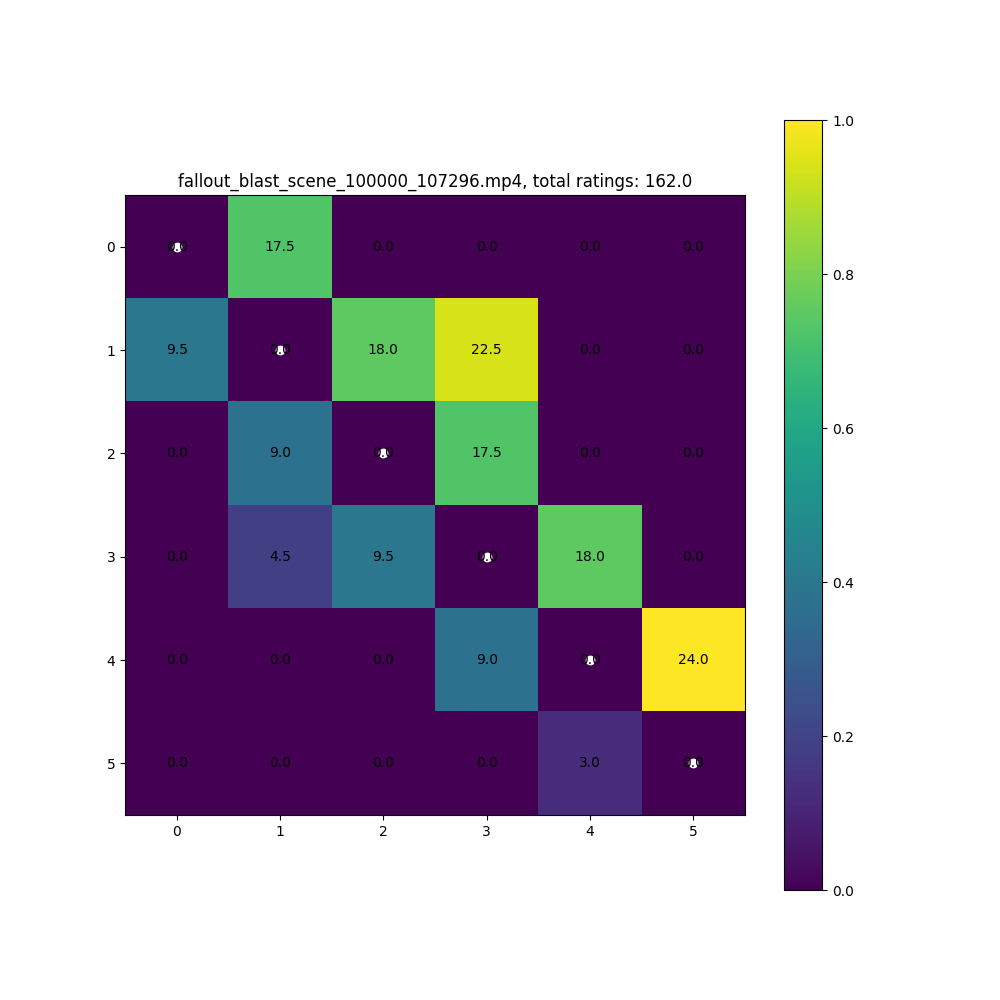}%
        \label{fig:falloutC_PCM}%
    }
    \caption{Recovered JOD scores (top) and PCM distributions (bottom) for a grainy blast scene across Groups A, B, and C.}
    \label{fig:fallout_triple}
\end{figure*}

Figure~\ref{fig:fallout_triple} shows recovered JOD scores and PCM heatmaps. Group~C participants made decisive distinctions with monotonic JOD progression, indicating feedback stabilized judgments. Group~A exhibited greater variability, reinforcing that training and feedback enhance perceptual quality differentiation, especially for subtle high-bitrate differences.

\section{Limitations and Discussion}

While our framework demonstrates substantial improvements in subjective video quality assessment, several limitations warrant discussion. Our study was conducted in a controlled laboratory setting with calibrated HDR10-capable OLED displays. The generalizability to diverse home viewing environments, where ambient lighting, display quality, and viewing distance vary significantly, remains to be validated through future crowdsourced studies. Additionally, while the controlled study reported here used HEVC (x265) encoding, the framework has been successfully deployed for AV2 codec evaluation and is codec-agnostic by design. Systematic validation across multiple codecs (AV1, VVC, VP9) with different artifact patterns remains future work to ensure the training materials and golden pair strategies generalize effectively.

Our participant pool of 80 lab-recruited subjects provides strong internal validity but limited external validity to broader populations with diverse viewing habits and quality expectations. The framework's effectiveness with different display technologies (LCD, LED, projection) and varied ambient conditions has not been systematically evaluated. Furthermore, our approach of splitting tie responses equally (50-50) in the PCM, while preserving mathematical properties under the Thurstone Case V model, could be compared with alternative models that explicitly represent ties as a third outcome category. The golden pair threshold parameters ($|\bar{r}| > 0.5$, $\sigma_r < 0.3$, 75\% agreement) and penalty/bonus function weights ($\Delta = 1.0 + 0.4 \times$ mistakes, $B = 1.0 + 0.2 \times$ correct) were empirically tuned based on our study population. While these parameters yielded robust results, formal sensitivity analysis across diverse populations and content types remains future work.

The chain-based pairwise comparison strategy was designed to achieve O(n) complexity comparable to ACR/DCR methods, making large-scale testing practical where exhaustive O(n²) pairwise comparisons would be prohibitively resource-intensive. For instance, 10 videos would require only 9 chain comparisons versus 45 full pairwise comparisons. This efficiency trade-off enables scalable quality assessment while recovering JOD scores that align with expected quality trends. Additionally, the current study focuses on compression artifacts inherent to video encoding (blockiness, banding, grain loss). However, the framework's training and attention scoring mechanisms can be readily extended to other quality dimensions such as frame drops, color depth reduction, dynamic range mapping errors, and transmission-related artifacts.

\section{Conclusions and Next Steps}
In this paper, we introduced a novel framework for subjective video quality assessment that combines automated participant training, real-time attention scoring, and an efficient pairwise comparison protocol to recover quality scales in JOD units. Our controlled study evaluated three participant groups—\emph{Group~A} (no training, no feedback), \emph{Group~B} (training without feedback), and \emph{Group~C} (training with real-time feedback)—and demonstrated that participant training significantly increases attentiveness and consistency in rating subtle quality differences. Real-time feedback further amplifies these benefits, as evidenced by \emph{Group~C}’s most monotonic JOD curves and highest confidence in distinguishing compression artifacts.

Additionally, by employing a chain-based comparison strategy, we reduced the number of pairwise comparisons from $\mathcal{O}(n^2)$ to $\mathcal{O}(n)$, thus lowering both cognitive load for participants and the overhead of test administration. Dynamically promoting “golden pairs” with strong consensus further refines attention scoring, reinforcing the identification of inattentive or random responses. Together, these design elements yield quality scores that align closely with expected bitrate trends while minimizing data contamination from inconsistent judgments.

The framework is deployed on a scalable cloud infrastructure, ensuring high availability and consistent video playback across devices, which seamlessly supports our training and attention scoring pipelines. Since deployment, the platform has collected over 27,000 ratings across 560 sessions, supporting critical Prime Video initiatives including AV1/AV2 codec standardization (with direct contributions to the AV2 proposal), validation of saliency-aware encoding models, and retraining of internal video quality metrics.

Future work will explore adaptive training modules that personalize content based on individual performance, crowdsourced validation across diverse populations and environments, extension to multiple codecs (AV1, VVC, VP9) for broader applicability, refined score recovery models that explicitly leverage tie statistics and golden pairs to better capture perceptual uncertainty, adaptive pair selection to reduce test duration, and integration with objective metrics (VMAF, SSIM) for hybrid quality models.

The authors would like to thank all participants and colleagues for volunteering for the subjective study.

{
    \small
    \bibliographystyle{ieee_fullname}
    \bibliography{references}

@misc{ITU_P910,
  title = {ITU-T P.910: Methods for the subjective assessment of the quality of television programmes},
  howpublished = {International Telecommunication Union},
  year = {1998}
}

@misc{VQEG,
  title = {VQEG Guidelines for the subjective assessment of video quality},
  howpublished = {Video Quality Experts Group},
  year = {2006}
}

@inproceedings{Mantiuk2012,
  author = {Mantiuk, R. K. and Tomaszewska, A. and others},
  title = {Comparison of four subjective methods for image quality assessment},
  booktitle = {Proceedings of the IEEE Conference on Image Processing},
  year = {2012},
  pages = {2478--2491}
}

@inproceedings{Ramanarayanan2007,
  author = {Ramanarayanan, G. and Ferwerda, J. and Walter, B.},
  title = {Visual equivalence: Towards a new standard for image fidelity},
  booktitle = {ACM Transactions on Graphics (TOG)},
  volume = {26},
  number = {3},
  pages = {76},
  year = {2007}
}

@misc{Mantiuk2017,
      title={A practical guide and software for analysing pairwise comparison experiments}, 
      author={Maria Perez-Ortiz and Rafal K. Mantiuk},
      year={2017},
      eprint={1712.03686},
      archivePrefix={arXiv},
      primaryClass={stat.AP},
      url={https://arxiv.org/abs/1712.03686}, 
}

@misc{E2NEST,
  title = {e2nest: A scalable platform for subjective media quality testing},
  howpublished = {\url{https://github.com/Netflix/e2nest}},
  year = {2020}
}

@inproceedings{goering2021voyager,
  title={AVRate Voyager: an open source online testing platform},
  author={Steve Göring and Rakesh {Rao Ramachandra Rao} and Stephan Fremerey and Alexander Raake},
  year={2021},
  booktitle={2021 IEEE 23st International Workshop on Multimedia Signal Processing (MMSP)},
  pages={1--6},
  organization={IEEE}
}

@misc{AVRateNG,
  author    = {AVRateNG},
  title     = {AVRateNG -- github project},
  howpublished = {\url{https://github.com/Telecommunication-Telemedia-Assessment/avrateNG}},
  year = {2024}
}

@INPROCEEDINGS{Figuerola2013,
  author={Figuerola Salas, Óscar and Adzic, Velibor and Kalva, Hari},
  booktitle={2013 Picture Coding Symposium (PCS)}, 
  title={Subjective quality evaluations using crowdsourcing}, 
  year={2013},
  volume={},
  number={},
  pages={418-421},
  doi={10.1109/PCS.2013.6737772}}

@inproceedings{Sakic2014,
  author={Šakić, Krešimir and Dumić, Emil and Grgić, Sonja},
  booktitle={IWSSIP 2014 Proceedings}, 
  title={Crowdsourced subjective Video Quality Assessment}, 
  year={2014},
  pages={223-226},
  url={https://ieeexplore.ieee.org/document/6837671}
}

@article{Burmania2016,
  author = {Alec Burmania and Carlos Busso and John Hansen},
  title = {Increasing the Reliability of Crowdsourcing Evaluations Using Online Quality Assessment},
  journal = {IEEE Transactions on Affective Computing},
  volume = {7},
  number = {4},
  pages = {374-388},
  year = {2016},
  doi = {10.1109/TAFFC.2015.2486743}
}

@INPROCEEDINGS{Egger2015,
  author={Hoßfeld, Tobias and Redi, Judith},
  booktitle={2015 Seventh International Workshop on Quality of Multimedia Experience (QoMEX)}, 
  title={Journey through the crowd: Best practices and recommendations for crowdsourced QoE}, 
  year={2015},
  volume={},
  number={},
  pages={1-2},
  doi={10.1109/QoMEX.2015.7148150}}

@ARTICLE{Orduna2023,
  author={Orduna, Marta and Pérez, Pablo and Gutiérrez, Jesús and García, Narciso},
  journal={IEEE Transactions on Affective Computing}, 
  title={Methodology to Assess Quality, Presence, Empathy, Attitude, and Attention in 360-degree Videos for Immersive Communications}, 
  year={2023},
  volume={14},
  number={3},
  pages={2375-2388},
  doi={10.1109/TAFFC.2022.3149162}}

@article{Mirkovic2014,
  author = {Mirkovic, M. and Vrgovic, P. and Culibrk, D. and Stefanovic, D. and Anderla, A.},
  title = {Evaluating the Role of Content in Subjective Video Quality Assessment},
  journal = {The Scientific World Journal},
  volume = {2014},
  pages = {625219},
  year = {2014},
  doi = {10.1155/2014/625219},
  pmid = {24523643},
  pmcid = {PMC3913198},
  url = {https://pmc.ncbi.nlm.nih.gov/articles/PMC3913198/}
}

@article{Kittel2023,
  author    = {Kittel, Anne Frieda Doris and Seufert, Tina},
  title     = {It's all metacognitive: The relationship between informal learning and self-regulated learning in the workplace},
  journal   = {PLOS ONE},
  year      = {2023},
  volume    = {18},
  number    = {5},
  pages     = {e0286065},
  doi       = {10.1371/journal.pone.0286065},
  url       = {https://journals.plos.org/plosone/article?id=10.1371/journal.pone.0286065}
}

@manual{x265,
  title = {x265: Open Source H.265/HEVC Video Encoder},
  author = {{MulticoreWare}},
  year = {2024},
  note = {Version 3.5, Available: \url{https://x265.readthedocs.io/en/stable/cli.html}}
}

@standard{HEVC,
  author = {{ITU-T and ISO/IEC}},
  title = {High Efficiency Video Coding},
  year = {2019},
  number = {ITU-T H.265 and ISO/IEC 23008-2},
  institution = {International Telecommunication Union and ISO/IEC},
  note = {Available: \url{https://www.itu.int/rec/T-REC-H.265}}
}

@INPROCEEDINGS{rahul_ICIP_2022,
  author={Zhu, Jingwen and Le Callet, Patrick and Perrin, Anne-Flore and Sethuraman, Sriram and Rahul, Kumar},
  booktitle={2022 IEEE International Conference on Image Processing (ICIP)}, 
  title={On The Benefit of Parameter-Driven Approaches for the Modeling and the Prediction of Satisfied User Ratio for Compressed Video}, 
  year={2022},
  volume={},
  number={},
  pages={4213-4217},
  doi={10.1109/ICIP46576.2022.9897946}}

@INPROCEEDINGS{rahul_ICIP_2023,
  author={Zhu, Jingwen and Ak, Ali and Le Callet, Patrick and Sethuraman, Sriram and Rahul, Kumar},
  booktitle={2023 IEEE International Conference on Image Processing (ICIP)}, 
  title={ZREC: Robust Recovery of Mean and Percentile Opinion Scores}, 
  year={2023},
  volume={},
  number={},
  pages={2630-2634},
  doi={10.1109/ICIP49359.2023.10222033}}

@inproceedings{rahul_ACM_2023,
author = {Zhu, Jingwen and Ak, Ali and Dormeval, Charles and Le Callet, Patrick and Rahul, Kumar and Sethuraman, Sriram},
title = {Subjective Test Environments: A Multifaceted Examination of Their Impact on Test Results},
year = {2023},
isbn = {9798400700286},
publisher = {Association for Computing Machinery},
address = {New York, NY, USA},
url = {https://doi.org/10.1145/3573381.3596470},
doi = {10.1145/3573381.3596470},
booktitle = {Proceedings of the 2023 ACM International Conference on Interactive Media Experiences},
pages = {298–302},
numpages = {5},
location = {Nantes, France},
series = {IMX '23}
}

@inproceedings{yixu_icme_2025,
  title={Video Quality Assessment for Resolution Cross-Over in Live Sports},
  author={Zhu, Jingwen and Chen, Yixu and Wei, Hai and Sethuraman, Sriram and Wu, Yongjun},
  booktitle={2025 IEEE International Conference on Multimedia and Expo (ICME)},
  year={2025},
  organization={IEEE}
}

@inproceedings{Naderi2024,
  author={Naderi, Babak and Cutler, Ross},
  booktitle={ICASSP 2024-2024 IEEE International Conference on Acoustics, Speech and Signal Processing (ICASSP)}, 
  title={A Crowdsourcing Approach to Video Quality Assessment}, 
  year={2024},
  pages={3380-3384},
  doi={10.1109/ICASSP48485.2024.10446509}
}

@inproceedings{Jenadeleh2024_feedback,
  author={Jenadeleh, Mohsen and Heß, Alexander and Del Pin, Saul Higuera and Ebrahimi, Touradj},
  booktitle={2024 16th International Conference on Quality of Multimedia Experience (QoMEX)}, 
  title={Impact of Feedback on Crowdsourced Visual Quality Assessment with Paired Comparisons}, 
  year={2024},
  pages={178-183},
  doi={10.1109/QoMEX61742.2024.10598256}
}

@article{Jenadeleh2024_jnd,
  author={Jenadeleh, Mohsen and Hamzaoui, Raouf and Reips, Ulf-Dietrich and Saupe, Dietmar},
  journal={IEEE Transactions on Circuits and Systems for Video Technology}, 
  title={Crowdsourced Estimation of Collective Just Noticeable Difference for Compressed Video With the Flicker Test and QUEST+}, 
  year={2024},
  volume={34},
  number={10},
  pages={9832-9845},
  doi={10.1109/TCSVT.2024.3399907}
}

@inproceedings{Keimel2012,
  author={Keimel, Christian and Habigt, Julian and Horch, Clemens and Diepold, Klaus},
  booktitle={2012 Picture Coding Symposium}, 
  title={Qualitycrowd — A framework for crowd-based quality evaluation}, 
  year={2012},
  pages={245-248},
  doi={10.1109/PCS.2012.6213338}
}

@inproceedings{Mohammadi2022,
  author={Mohammadi, Saeed and Ascenso, João},
  booktitle={2022 IEEE International Conference on Image Processing (ICIP)}, 
  title={Evaluation of Sampling Algorithms for a Pairwise Subjective Assessment Methodology}, 
  year={2022},
  pages={2356-2360},
  doi={10.1109/ICIP46576.2022.9897942}
}
}

\end{document}